%% file: ms.tex
\documentclass[orivec]{llncs}
\pdfoutput=1
\usepackage{etex}
\usepackage[utf8]{inputenc}
\usepackage[T1]{fontenc}
\usepackage{amsmath,stmaryrd,mathpartir}
\usepackage{amssymb}
\usepackage{proof}
\usepackage[]{graphicx}
\usepackage{paralist}
\usepackage{enumitem}
\usepackage{xspace}
\usepackage{url}
\usepackage{bussproofs}
\usepackage[inference]{semantic}
\usepackage{fancyhdr}
\usepackage{epsfig}
\usepackage{latexsym}
\usepackage{lmodern}
\usepackage{microtype}
\usepackage{lastpage} 
\usepackage{hyperref}
\usepackage{soul}

\usepackage{listings}
\lstdefinestyle{customhaskell}{
    language=Haskell,
    showstringspaces=false,
    basicstyle=\footnotesize\rmfamily,
    keywordstyle=\bfseries\color{green!40!black},
    commentstyle=\itshape\color{blue!40!black},
    identifierstyle=\color{black},
    stringstyle=\color{orange!40!black},
}
\usepackage[dvipsnames*,svgnames]{xcolor}
\hypersetup{colorlinks=true,allcolors=DarkBlue,linkcolor=black}

\usepackage{tikz}
\usetikzlibrary{decorations.text, positioning}
\usetikzlibrary{arrows.meta}

\usepackage{z-eves}

\input{macros}

\begin{document}
\pagestyle{headings}
\bibliographystyle{plain}
%
\title{Towards a formally verified implementation of the MimbleWimble cryptocurrency protocol}
\author{Gustavo Betarte\inst{1} \and Maximiliano Cristi\'a\inst{2} \and Carlos Luna\inst{1} \and Adri\'an Silveira\inst{1} \and Dante Zanarini\inst{2}} 
\institute{InCo, Facultad de Ingenier\'ia, Universidad de la Rep\'ublica, Uruguay. 
   \\ \email{\{gustun,cluna,adrians\}@fing.edu.uy} 
\and CIFASIS, Universidad Nacional de Rosario, Argentina.
 \\ \email{\{cristia,zanarini\}@cifasis-conicet.gov.ar}
}

\maketitle

\begin{abstract}                                          
\input{abstract}
\end{abstract}
\input{introduction}
\input{background}
\input{model}
\input{properties}
\input{conclusion}
\bibliography{ms}  

\newpage
\appendix

\input{zModel}

\input{setlogModel}

\end{document}

%% file: macros.tex
\newcommand{\eqdef}{\stackrel{{\rm def}}{=}}

\newcommand{\step}[1]{\mathbin{\lower0.55ex\hbox{$\lhook\joinrel\xrightarrow{#1}$}}}

\newlength{\bcextramargin}
\newenvironment{changemargin}[2]{\begin{list}{}{%
\setlength{\topsep}{0pt}%
\setlength{\leftmargin}{0pt}%
\setlength{\rightmargin}{0pt}%
\setlength{\listparindent}{\parindent}%
\setlength{\itemindent}{\parindent}%
\setlength{\parsep}{0pt plus 1pt}%
\addtolength{\leftmargin}{#1}%
\addtolength{\rightmargin}{#2}%
}\item }{\end{list}} 

\newcommand{\actdefsection}[1]{
\begin{changemargin}{-\bcextramargin}{0pt}
\vspace{1ex}
\noindent
\textbf{{#1}}
\end{changemargin}
}

\newenvironment{absolutelynopagebreak}
  {\par\nobreak\vfil\penalty0\vfilneg
   \vtop\bgroup}
  {\par\xdef\tpd{\the\prevdepth}\egroup
   \prevdepth=\tpd}

\newcommand{\setlog}{$\{log\}$\xspace}

%% file: abstract.tex
%
MimbleWimble is a  privacy-oriented cryptocurrency technology encompassing  security and scalability properties that distinguish it from other protocols of the kind.
In this paper we present and briefly discuss those properties and outline the basis of a model-driven verification approach to address the certification of the correctness of a particular  implementation of  the protocol.

%% file: introduction.tex
\section{Introduction}
\label{sec:intro}


Cryptocurrency protocols deal with virtual money so they are a valuable target of highly skilled attackers. Some attacks have already been mounted against cryptocurrency systems causing irreparable loses of money and credibility (e.g. \cite{dao}). For this reason the cryptocurrency community is seeking for approaches, methods, techniques and development practices that can reduce the chances of successful attacks. One such approach is the application of formal methods to software implementation. In particular, the cryptocurrency community is showing interest in formal proof and formally certified implementations. 

MimbleWimble is a privacy-oriented cryptocurrency technology encompassing  security and scalability properties that distinguish it from other technologies of the kind. Mimblewimble differs from Bitcoin \cite{bitcoin}, for  instance, in key areas: there is no concept of address and all the transactions are confidential. In this (short) paper we outline an approach based on formal software verification aimed at formally verifying the basic mechanisms of MimbleWimble and (one of) its implementations \cite{mimbleimble-doc}.


Reasoning about implementations provides the ultimate guarantee that deployed critical systems provide the expected properties. There are however significant hurdles with this approach.
Formally proving non-trivial security properties of code might be an overwhelming task in terms of the effort required, especially compared with proving functional correctness. In addition, many implementation details are orthogonal to the security properties to be established. This implies that slight changes in the implementation technology might have devastating consequences as concerns the security of the implementation. Therefore, complementary approaches are needed when non-trivial security properties are at stake. In this paper we put forward a model-driven verification approach where in the first place security issues that pertain to the realm of critical mechanisms are explored on an idealized model of the system under study. One such model abstracts away the specifics of any particular implementation and yet provides a realistic setting. Then verification is performed on more concrete models where low level mechanisms (such as pointer arithmetic) are specified; and finally the low level model is proved to be a correct implementation of the idealized model.

\paragraph{{\bf Organization of the paper}}
Section~\ref{sec:background} provides a very brief description of the MimbleWimble cryptocurrency protocol. 
Section \ref{sec:model} describes the building blocks of a formal idealized model (abstract state machine) of the computational behaviour of MimbleWimble and Section~\ref{sec:verification-MW} provides a brief account of the verification activities we are putting in place in order to verify the protocol and its implementation.
Final remarks are presented in Section \ref{sec:conclusion}.

%% file: background.tex
\section{The MimbleWimble protocol} \label{sec:background}
\label{MW}
%
Transactions are at the core of the Mimblewimble protocol and they constitute a derivation of what are known as confidential transactions \cite{GMaxwell:conftransac,AGibson:conftransac}. 

A confidential transaction allows a sender to encrypt the amount of bitcoins he wants to send by using blinding factors, which are values chosen by the sender that are  used to encrypt bitcoin amounts in a transaction. In a confidential transaction only the two parties involved know the amount of bitcoins being transacted. However, onlookers can still ensure that the transaction is valid by comparing the number of inputs and outputs; if both are the same, then the transaction will be considered valid. Such a procedure ensures that no bitcoins have been created from nothing and is key in preserving the integrity of the system.

Mimblewimble transactions function in a similar way, except  the recipient of a transaction randomly selects a range of blinding factors provided by the sender. This blinding factor is then used as proof of ownership by the receiver, thus, permitting him to spend the bitcoins.

The MimbleWimble protocol aims at providing the following properties \cite{mimbleimble-wp,mimbleimble-doc}:
\begin{inparaenum}[i)] 
\item verification of zero sums without revealing the actual amounts involved in a transaction, which implies anonymity;
\item authentication of transaction outputs without signing the transaction; and 
\item good scalability, while preserving security, by generating blocks of smaller sizes---or better, the size of old blocks can be reduced thus producing a blockchain whose size does not grow in time as much as, for instance, Bitcoin's.
\end{inparaenum} 
%
%
%

The first two properties are achieved by resting, in the end, on Elliptic Curves Cryptography (ECC) operations and properties. 
The third one is a consequence of the first two.

\paragraph{{\bf Verification of transactions}}
If $v$ is the value of a transaction (either input or output) and $H$ is a point over an elliptic curve, 
then $vH$ encrypts $v$ because it is assumed to be computationally hard to get $v$ from $vH$ if we only know $H$. However, if $w$ and $z$ are other values such that $v + w = z$, then if we only have the result of encrypting each of them with $H$ we are still able to verify that equation. Indeed:
$
v + w = z \iff vH + wH = zH,
$
due to simple properties of scalar multiplication over groups. Therefore, with this simple operations, we can check sums of transactions amounts without knowing the actual amounts. 

Nevertheless, say some time ago we have encrypted $v$ with $H$ and now we see $vH$, then we know that it is the result of encrypting $v$. In the context of blockchain transactions this is a problem because once a block holding $vH$ is saved in the chain it will reveal all the transactions of $v$ coins. For such problems, MimbleWimble encrypts $v$ as $rG + vH$ where $r$ is an scalar and $G$ is another point in $H$'s elliptic curve. $r$ is called \emph{blinding factor} and $rG + vH$ is called \emph{Pedersen Commitment}. By using Pedersen Commitments, MimbleWimble allows for verification of expressions such as $v + w = z$ providing more privacy than with the standard scheme. In effect, if $v + w = z$ then we chose $r_v$, $r_w$ and $r_z$ such that $r_vG + r_wG = r_zG$ and so the expression is recorded as:
$$
\overbrace{(r_vG + vH)}^v + \overbrace{(r_wG + wH)}^w = 
\overbrace{r_zG + zH}^z
$$
making it possible for every one to verify the transaction without knowing the true values.

\paragraph{{\bf Authentication of transactions}}
Consider that Alice has received $v$ coins and this was recorded somewhere in the blockchain as $rG + vH$, where $r$ was chosen by her to keep it private. Now she wants to transfer these $v$ coins to Bob. As a consequence, Alice looses $v$ coins and Bob receives the very same amount, which means that the transaction adds to zero: $rG + vH - (rG + vH) = 0G - 0H$. However, Alice now knows Bob's blinding factor because it must be the same chosen by her (so the transaction is balanced). In order to protect Bob from being stolen by Alice, MimbleWimble allows Bob to add his blinding factor, $r_B$, in such a way that the transaction is recorded as:
$
(r + r_B)G + vH - (rG + vH) = r_BG - 0H,
$
although now it does not sum zero. However, this \emph{excess value} is used as part of an authentication scheme. Indeed, Bob uses $r_B$ as a private key to sing, say, the empty string ($\epsilon$). This signed document is attached to the transaction so in the blockchain we have:
\begin{inparaenum}[1)] 
\item Input: $I$; 
\item Output: $O$; and 
\item Bob's signed document: $S$.
\end{inparaenum} 
In this way, in order to verify the transaction one has to see whether the result of decrypting $S$ with $I - O$ (in the group generated by $G$) yields $\epsilon$. If $I - O$ does not yield something of the form $r_BG - 0H$, then $\epsilon$ will not be recovered and so we know there is an attempt to create money from thin air or there is an attempt to still Bob's money.
%

%% file: model.tex
\section{Idealized model of a Mimblewimble-based blockchain}
\label{sec:model}

The basic elements in our model are transactions, blocks and chains. Each node in the blockchain maintains a local state. The main components are the local copy of the chain and the set of transactions waiting to be validated and added to a new block. Properties as zero-sum and the absence of double spending in blocks and chains must be proved for local states. 

The blockchain global state can be represented as a mapping from nodes to local states. For global states, we can state and prove properties of the entire system like, for instance,  correctness of the consensus protocol. 

\paragraph{{\bf Transactions}}
Given two elliptic curves $G$ and $H$, we represent transactions as tuples of the form:
$$
\begin{array}{lcl}
 Transaction &\eqdef& \{i : I^{\star},\ o : O^{\star}, tk : TxKernel^{\star} \}
\end{array}
$$ where $X^{\star}$ represents the lists of $X$ elements.
Each element in $i$ and $o$ is a pair of the form $(r, v)$
representing the Pedersen commitment in curves $G$, $H$.  The transaction kernel $tk$ contains a list of range proofs of the outputs, a list of transaction excess and the kernel signature $\sigma$\footnote{For simplicity, we leave aside fees in this paper.}.
The transaction excess can be defined as: 
$$ ke = \sum_{(r',v') \in o}{r'.G + v'.H} - \sum_{(r,v) \in i}{ r.G + v.H}\ $$
We said that the transaction is balanced iff $\sum_{}{r'} - \sum_{}{r} = 0$. So, in other words, the excess is $(\sum_{}{v'} - \sum_{}{v})H$.
The kernel signature proves that the transaction is honestly constructed, in particular that the excess only contains the blinding factor (no money is being created).

A {\it transaction is valid} if:
\begin{inparaenum}[1)] 
\item the range proofs of all the outputs are valid;
\item the kernel signature $\sigma$ is valid for the excess; and
\item the transaction is balanced.
\end{inparaenum}	

\paragraph{{\bf Blocks and chains}}
Transactions can be merged into a {\it block}. We can see a block as a big transaction with aggregated inputs, outputs and transaction kernels.

A {\it Block} is either the {\it genesis block $\mathit{Gen}$}, or  a record: 
$$
\begin{array}{lcl}
 Block &\eqdef& \{i : I^{\star},\ o : O^{\star},\ k : \nat ,\ tks : TxKernel^{\star}\} \\
\end{array}
$$
representing the inputs, outputs, the kernel offset and the list of transaction kernels for the block. The kernel offset $k$ is a blinding factor that needs to be added back to the kernel excess to verify the commitments sum to zero.


Given a block $b = \{i,o,k,tks\}$, we say that a block is valid if the following equality holds:

$$ \sum_{(r',v') \in o}{(r'.G+v'.H)} - \sum_{(r,v) \in i}{(r.G+v.H)}
    = 
    k + \sum_{(ke,s) \in tks}{ke}$$
    
A {\it chain} is a non-empty list of {\it blocks}:
$$\mathit{Chain} \eqdef \mathit{Block}^{\star}$$

For a chain $c$ and a valid block $b$, we can define a predicate 
$\mathit{validates(c, b)}$ representing the fact that is valid to add $b$ to $c$. This relation must check, for example, that all the inputs in $b$ are present as outputs in $c$, in other words, they are unspent transaction outputs (UTXOs). 
    
\paragraph{{\bf Validating a chain}}
The model formalizes a notion of valid state that captures several well-formedness conditions. In particular, every block in the blockchain must be valid.
A predicate $\mathit{validChain}$ can be defined for a chain 
$c = [b_{0}, b_{1}, \dots b_{n}]$ by checking that:
\begin{inparaenum}[1)] 
 \item $b_0$ is a valid genesis block, and
 \item for every $i\in \{1, \dots n\}$, 
 $\mathit{validates([b_{0}, \dots, b_{i-1}], b_{i})}.$
\end{inparaenum}
%

The axiomatic semantics of the system is modeled by defining a set of actions, and providing their semantics as state transformers.  
The behaviour of actions is specified by a precondition and by a postcondition. This approach is valid when considering local (nodes) or global (blockchain) states and actions. Different set of actions, pre and postcondition are defined to cover local or global state transformations.

Valid states are invariant under execution. The properties in this work are obtained from valid states of the system.

%

%% file: properties.tex
\newtheorem{prop}{Property}

\section{Verification of MimbleWimble}
\label{sec:verification-MW}
We now proceed to discuss some relevant properties that can be verified on our model.
In addition to some of the properties mentioned in Section \ref{MW} we also plan to include other properties such as those formulated in \cite{Pirlea:2018:MBC:3176245.3167086}, and various security properties considered in \cite{DBLP:conf/eurocrypt/GarayKL15,DBLP:conf/crypto/KiayiasRDO17,FuchsbauerOS19} later.

\subsubsection{Protocol Properties.} 
\label{sec:properties}
The \textit{zero-sum} for valid transactions can be proved using properties of ECC. This property, and its proof, straightforwardly propagates to blocks, because adding zero-sum Pedersen commitments is a zero-sum Pedersen commitment. 

An important feature of MimbleWimble is the \textit{cut-through} process. The purpose of this property is to erase redundant outputs that are used as inputs within the same block. Let $C$ be some coins that appear as an output in the block $b$. If the same coins appear as an input within the block, then $C$ can be removed from the list of inputs and outputs after applying the  cut-through process. In this way, eventually, the only data that remain are the block headers, transaction kernels and unspent transaction outputs (UTXOs).
After applying cut-through to a valid block $b$ is important to ensure that the resulting block $b'$ is still valid. We can say that the validity of a block should be invariant with respect to the cut-through process. Basically, this invariant holds because the matching inputs and outputs canceled each other in the overall sum. 
\paragraph{Privacy and Security Properties.} 
\label{sec:sec_props}
In blockchain systems the notion of privacy is crucial: sensitive data should not be revealed over the network. In particular,  it is desirable to ensure properties such as  confidentiality, anonymity and unlinkability of transactions. The first one refers to the property of preventing other participants from knowing certain information about the transaction, such as the amounts and addresses of the owners. The second one refers to the property of hiding the real identity from the one is transacting while the third one refers to the inability of linking distinct transactions of the same user within the blockchain. 

In  the Bitcoin network, users interact with the system using public key hashes achieving a kind of pseudo-anonymity. Every transaction recorded on the ledger contains the addresses of the sender and receiver and the transaction amount. That information is publicly available  breaking then  the confidentiality property. Furthermore, it fails in providing unlinkability since it is possible to trace the transactions of the same associated addresses over the ledger. 

In the case of Mimblewimble no addresses or public keys are used, there are only encrypted inputs and outputs. Therefore the communication between the sender and receiver to share the proof of ownership of the coins must be done off-chain using a secure channel. Privacy concerns rely on confidential transactions, cut-through and CoinJoin. The goal of CoinJoin is to combine inputs and outputs from different transactions into a single unified transaction. Thus, for a third-party it is difficult to determine which party is making a particular transaction. It is important to ensure that the resulting transaction satisfies the property of validity defined in the model.

The security problem of double spending refers to spend a coin more than once. All the nodes keep track of the UTXO set, so before confirming a block to the chain, the node checks that the inputs come from it. If we refer to our model, it is performed in the predicate \textit{validates} mentioned in Section \ref{sec:model}.  

\paragraph{Zero-knowledge Proof.}
The goal of this kind of proofs is to prove that a statment is true without revealing any information beyond the verification of the statment. These proofs could require interaction between the prover and the verifier or not. 
Genereally, when there is no interaction the validity of the proof relies on computational assumptions. 
In MimbleWimble  we need to ensure that in every transaction the transaction amount is positive so that users cannot create coins. The key here is to prove that property without revealing the amount. As we defined in the model, the output amounts are hidden in the form of a Pedersen Commitment and the transaction contains a list of range proofs of the outputs to prove that the transaction amount is positive. MimbleWimble uses Bulletproofs \cite{BunzBullet} to achieve this goal\footnote{This proof rely on the discrete logarithm assumption and a trusted setup is not required. The proof size is logarithmic in the amount size and the proof generation and verification are linear in the bit length of the range.}. In our model, this verification is performed as first step of the validity of the transaction. 

\subsubsection{\bf Model-driven verification.}
MimbleWimble is built on top of a consensus protocol. 
In that direction, we have developed a Z specification of a consensus protocol (see Appendix \ref{zModel}). Z specifications in turn can be easily translated into the \setlog language \cite{Cristia2019}. \setlog can be used as both a (prototyping) programming language and an automated theorem prover for an expressive fragment of set theory and set relation algebra. We include an excerpt of the \setlog prototype of a consensus protocol in Appendix \ref{setlogModel}.  This \setlog prototype can be used as an executable model where simulations can be run. This allows us to analyze the behavior of the protocol without having to implement it in a low level programming language.
%

We also plan to use \setlog to prove some of the properties mentioned above.  However, for complex properties or for properties not expressible in the set theories supported by \setlog we plan to develop a complete and uniform formulation of several security properties of the protocol using the \texttt{Coq} proof assistant \cite{coq-manual}. \texttt{Coq} has an important set of libraries; for example \cite{DBLP:conf/itp/BartziaS14} contains a formalization of elliptic curves theory, which allows the verification of elliptic curve cryptographic algorithms. 
The fact of first having a \setlog prototype over which some verification activities can be carried out without much effort helps in simplifying the process of writing a detailed \texttt{Coq} specification. This is in acordance with proposals such as \texttt{QuickChick} whose goal is to decrease the number of failed proof attempts in \texttt{Coq} by generating counterexamples before a proof is attempted \cite{denes2014quickchick}.

Applying the program extraction mechanism provided by \texttt{Coq} we shall be able  to derive a certified Haskell prototype of the protocol, which can be used as a testing oracle and also to conduct further verification activities on correct-by-construction implementations of the protocol. In particular, both the \setlog and \texttt{Coq} approaches can be used as forms of model-based testing. That is, we can use either specification to automatically generate test cases with which protocol implementations can be tested \cite{CristiaRossiSEFM13,denes2014quickchick}.

%% file: conclusion.tex
\section{Final remarks}
\label{sec:conclusion}
We have put forward elements that constitute essential steps towards the development of an exhaustive formalization  of the MimbleWimble cryptocurrency protocol, the analysis of its properties  and the verification of its implementations. In particular, the proposed idealized model is key in the described verification process.


%% file: zModel.tex
\section{\label{zModel}Excerpt of a Z Model of a Consensus Protocol}
The following are some snippets of a Z model of a consensus protocol based on the model developed by P\^irlea and Sergey \cite{Pirlea:2018:MBC:3176245.3167086}. For reasons of space we just reproduce a little part of it.

The time stamps used in the protocol are modeled as natural numbers. Then we have the type of addresses ($Addr$), the type of hashes ($Hash$), the type of proofs objects ($Proof$) and the type of transactions ($Tx$). Differently from P\^irlea and Sergey's model\footnote{From now on we will refer to P\^irlea and Sergey model simply as PS.} we modeled addresses as a given type instead as natural numbers. In PS the only condition required for these types is that they come equipped with equality, which is the case in Z.
\begin{zed}
Time == \nat \also
[Addr,Hash,Proof,Tx]
\end{zed}
The block data structure is a record with three fields: $prev$, (usually) points to the parent block; $txs$, stores the sequence of transactions stored in the block; and $pf$ is a proof object required to validate the block.
\begin{schema}{Block}
prev:Hash \\
txs: \seq Tx \\
pf:Proof
\end{schema}
The local state space of a participating network node is given by three state variables: $as$, are the addresses of the peers this node is aware of; $bf$, is a block forest (not shown) which records the minted and received blocks; and $tp$, is a set of received transactions which eventually will be included in minted blocks.
\begin{schema}{LocState}
as:\power Addr \\
bf: Hash \pfun Block \\
tp: \power Tx
\end{schema}
The system configuration is represented by two state variables: $Delta$, which establishes a mapping between network addresses and the corresponding node (local) states (in PS this variable is referred to as the \emph{global state}); and $P$, a set of packets (which represent the messages exchanged by nodes). 
\begin{schema}{Conf}
Delta:Addr \pfun LocState \\
P:\power Packet
\end{schema}
Packets are just tuples of two addresses (origin and destination) and a message.
\begin{zed}
Packet == Addr \cross Addr \cross Msg
\end{zed}
The model has twelve state transitions divided into two groups: \emph{local} and \emph{global}. Local transitions are those executed by network nodes, while global transitions promote local transitions to the network level. In turn, the local transitions are grouped into \emph{receiving} and \emph{internal} transitions. Receiving transitions model the nodes receiving messages from other nodes and, possibly, sending out new messages; internal transitions model the execution of instructions run by each node when some local condition is met. Here, we show only the local, receiving transition named $RcvAddr$.
\begin{schema}{RcvAddr}
\Delta LocState \\
p?:Packet \\
ps!:\power Packet
\where
p?.2 = this \\
\exists asm:\power Addr @ \\
   \t1 p?.3 = AddrMsg~asm \\
   \t1 \land as' = as \cup asm \\
   \t1 \land bf' = bf \\
   \t1 \land tp' = tp \\
   \t1 \land ps! = \{a:asm \setminus as @ (p?.2,a,ConnectMsg)\}~\cup \\
        \t1 ~~~~~\{a:as @ (p?.2,a,AddrMsg~as')\}
\end{schema}
As can be seen, $RcvAddr$ receives a packet ($p?$) and sends out a set of packets ($ps!$). The node checks whether or not the packet's destination address coincides with its own address. In that case, the node adds the received addresses to its local state and sends out a set of packets that are either of the form $(p?.2,a,ConnectMsg)$ or $(p?.2,a,AddrMsg~as')$. The former are packets generated from the received addresses and sent to the new peers the node now knows, while the latter are messages telling its already known peers that it has learned of new peers.

%% file: setlogModel.tex
\section{\label{setlogModel}Excerpt of a \setlog Prototype of a Consensus Protocol}
In this section we show the \setlog code corresponding to the Z model presented in Appendix \ref{zModel}. \setlog code can be seen as both a formula and a program \cite{Cristia2019}. Thus, in this case we use the code as a prototype or executable model of the Z model. The intention is twofold: to show that passing from a Z specification to a \setlog program is rather easy, and to show how a \setlog program can be used as a prototype. The first point is achieved mainly because \setlog provides the usual Boolean conectives and most of the set and relational operators available in Z. Hence, it is quite natural to encode a Z specification as a \setlog program.

Given that \setlog is based on Prolog its programs resemble Prolog programs. The \setlog encoding of $RcvAddr$ is the following:
\begin{verbatim}
rcvAddr(LocState,P,Ps,LocState_) :-
LocState = {[as,As] / Rest} &
P = [_,this, addrMsg(Asm)] &
un(As,Asm,As_) &
diff(Asm,As,D) &
Ps1 = ris(A in D,[],true,[this,A,connectMsg]) &
Ps2 = ris(A in As,[],true,[this,A,addrMsg(As_)]) &
un(Ps1,Ps2,Ps) &
LocState_ = {[as,As_] / Rest}.
\end{verbatim}

As can be seen, \verb+rcvAddr+ is clause receiving the before state (\verb+LocState+), the input variable (\verb+P+), the output variable (\verb+Ps+) and the after state (\verb+LocState_+). As in Prolog, \setlog programs are based on unification with the addition of set unification. In this sense, a statement such as \verb+LocState = {[as,As] / Rest}+ (set) unifies the parameter received with a set term singling out the state variable needed in this case (\verb+As+) and the rest of the variables (\verb+Rest+). The same is done with packet \verb+P+ where \verb+_+ means any value as first component and \verb+addrMsg(Asm)+ gets the set of addresses received in the packet without introducing an existential quantifier.

The set comprehensions used in the Z specification are implemented with \setlog's so-called Restricted Intentional Sets (RIS) \cite{DBLP:conf/cade/CristiaR17}. A RIS is interpreted as a set comprehension where the control variable ranges over a finite set (\verb+D+ and \verb+As+). 

Given \verb+rcvAddr+ we can perform simulations on \setlog such as:
\begin{verbatim}
S = {[as,{}] / R} & 
rcvAddr(S,[_,this,addrMsg({a1,a2})],P1,S1) & 
rcvAddr(S1,[_,this,addrMsg({a1,a3})],P2,S2).
\end{verbatim}
in which case \setlog returns:
\begin{verbatim}
P1 = ris(A in {a1,a2/_N2},[],true,[this,A,connectMsg],true),  
S1 = {[as,{a1,a2}]/R},  
P2 = {[this,a3,connectMsg],[this,a1,addrMsg({a2,a1,a3})],
      [this,a2,addrMsg({a2,a1,a3})] /
      ris(A in _N1,[],true,[this,A,connectMsg],true)},  
S2 = {[as,{a2,a1,a3}]/R}
Constraint: subset(_N2,{a1,a2}), subset(_N1,{a1,a3}), 
            a1 nin _N1, a2 nin _N1
\end{verbatim}
That is, \setlog binds values for all the free variables in a way that the formula is satisfied (if it is satisfiable at all). In this way we can trace the execution of the protocol w.r.t. states and outputs by starting from a given state (e.g. \verb+S+) and input values (e.g. \verb+[_,this,addrMsg({a1,a2})]+), and chaining states throughout the execution of the state transitions included in the simulation (e.g. \verb+S1+ and \verb+S2+).